\documentclass[fleqn,usenatbib]{mnras}

\usepackage{mathptmx}

\usepackage[T1]{fontenc}
\usepackage{ae,aecompl}

\usepackage{graphicx}	
\usepackage{amsmath}	
\usepackage{amssymb}	
\usepackage{multicol}   
\usepackage{bm}		    
\usepackage{pdflscape}	
\usepackage{epsfig,color,pifont}
\usepackage{times}
\usepackage{subfig}
\usepackage{url}

\usepackage{ulem}

\usepackage{graphicx}	
\usepackage{amsmath}	
\usepackage{amssymb}	

\def\diff{{\rm d}}
\def\Msun{\mbox{~M$_\odot$}}
\def\Lsun{\mbox{~L$_\odot$}}

\def\Msunpc2{\mbox{~M$_\odot$~pc$^{-2}$}}
\def\kms{\mbox{~km~s$^{-1}$}}

\def\pc{\mbox{~pc}}
\def\kpc{\mbox{~kpc}}
\def\Mpc{\mbox{~Mpc}}

\def\Myr{\mbox{~Myr}}
\def\Gyr{\mbox{~Gyr}}
\def\erg{\mbox{~erg}}

\def\tsnO{t_{\rm OB,0}}

\def\epscore{\epsilon_{\rm core}}

\def\epsff{\epsilon_{\rm ff}}

\def\SigmaISM{\Sigma_{\rm ISM}}

\def\Mmin{M_{\rm min}}
\def\MToomre{M_{\rm Toomre}}
\def\Mmax{M_{\rm max}}

\def\vcirc{V_{\rm circ}}

\def\Vc{V_{\rm circ}}
\def\Vcirc{V_{\rm circ}}

\def\Rvir{R_{\rm vir}}
\def\R200{R_{200}}
\def\rhos{\rho_{\rm s}}
\def\rhocrit{\rho_{\rm crit}}
\def\rs{r_{\rm s}}
\def\rcore{r_{\rm core}}
\def\Mvir{M_{\rm vir}}
\def\Mtot{M_{\rm tot}}

\def\Mstar{M_*}
\def\Mhalo{M_{\rm halo}}

\def\rs{r_{\rm s}}
\def\apj{\rm Astrophys. J.}
\def\aap{\rm Astron. \&Astrophys. }
\def\mnras{\rm Mon.\,Not.\,RAS. }

\def\Msunh{~h^{-1}{\rm M}_\odot}

\def\mathnew{\mathsurround=0pt}
\def\simov#1#2{\lower .5pt\vbox{\baselineskip0pt
    \lineskip-.5pt\ialign{$\mathnew#1\hfil##\hfil$\crcr#2\crcr\sim\crcr}}}

\def\lesssim{\mathrel{\mathpalette\simov <}}
\def\apj{\rm ApJ}
\def\apjl{\rm ApJL}
\def\apjs{\rm ApJS}
\def\aj{\rm AJ}
\def\mnras{\rm MNRAS}
\def\nat{\rm Nature}

\def\apss{\rm APSS}
\def\aap{\rm A\&A}
\def\araa{\rm ARAA}

\newcommand{\HI}{\hbox{{\sc H}\hspace{0.7pt}{\sc i}} }

\defcitealias{Reina-Campos17}{RK17}
\defcitealias{Kruijssen12d}{K12}
\defcitealias{Trujillo-Gomez19}{TRK19}

\title[GCs and the formation of DF2]{Constraining the formation of NGC1052-DF2 from its unusual globular cluster population}

\author[S. Trujillo-Gomez et al.]
{Sebastian Trujillo-Gomez\thanks{E-mail: strujill@gmail.com}, 
J.~M.~Diederik Kruijssen, 
Benjamin W.~Keller,
\newauthor
 and Marta Reina-Campos
\\
Astronomisches Rechen-Institut, Zentrum f{\"u}r Astronomie der Universit{\"a}t Heidelberg, Monchhofstra{\ss}e 12-14, D-69120 Heidelberg, Germany
}

\date{Accepted 2021 June 17. Received 2021 June 17; in original form 2020 September 9}

\pubyear{2020}

\begin{document}
\label{firstpage}
\pagerange{\pageref{firstpage}--\pageref{lastpage}}
\maketitle

\begin{abstract}
The ultra-diffuse galaxy (UDG) NGC1052-DF2 has a low dark matter content and hosts a very unusual globular cluster (GC) population, with a median luminosity $\sim4$ times higher than in most galaxies and containing about 5~per~cent of the galaxy's stars. We apply a theoretical model that predicts the initial cluster mass function as a function of the galactic environment to investigate the origin of DF2's peculiar GC system. Using the GC mass function, the model constrains the star-forming conditions in the galaxy during the formation of its GCs, $\sim9\Gyr$ ago. We predict that the GCs formed in an environment with very high gas surface density, $\SigmaISM \ga 10^3\Msun\pc^{-2}$, and strong centrifugal support, $\Omega \ga 0.7\Myr^{-1}$, similar to  nearby circum-nuclear starbursts and the central region of the Milky Way. The extreme conditions required to form the observed GC population imply a very high cluster formation efficiency of $\approx 87$~per~cent, and contrast strongly with the current diffuse nature of the galaxy. Since a nuclear starburst would lead to the rapid in-spiral of the GCs and is ruled out by the absence of a nuclear star cluster, we propose that the GCs plausibly formed during a major merger at $z\sim1.3$. The merger remnant must have undergone significant expansion of its stellar (and perhaps also its dark matter) component to reach its low present surface brightness, leading to the interesting possibility that it was the formation of DF2's extreme GC population that caused it to become a UDG. If true, this strong structural evolution would have important implications for understanding the origins of UDGs.
\end{abstract}




\section{Introduction}
\label{sec:intro}

Ultra-diffuse galaxies (UDGs) are galaxies with effective radii as large as those of $L^*$ galaxies\footnote{Although the truncation of their stellar component occurs at a similar radius as in dwarf galaxies \citep{Chamba20}.}, but with a considerably lower luminosity and surface brightness. They have been found preferentially in galaxy clusters \citep{Caldwell06,vanDokkum15, Koda15, Mihos15, vandenBurg16, Mancera-Pina19a}, but also in galaxy groups \citep{Merritt16,Trujillo17,RomanTrujillo17b}, around MW-mass galaxies \citep{Henkel17}, and in the field \citep{Leisman17,Mancera-Pina19b}. They appear red and spheroidal when found inside galaxy clusters \citep{vanDokkum15}, and typically \HI-rich with ongoing star formation and irregular morphologies when found in isolation \citep{Prole19}. Although their properties and origin are still quite uncertain, they seem to inhabit dark matter (DM) haloes with a broad range of masses, with some UDGs hosted by haloes as massive as the MW ($\sim 10^{12}\Msun$) \citep{vanDokkum16,vanDokkum17}, while most have either spatial distributions, mass functions, total disc extent, or GC properties (numbers, colours, or kinematics) that suggest their DM haloes are comparable in mass to those of dwarf galaxies, $\sim 10^{10} - 10^{11}\Msun$ \citep{Beasley16,BeasleyTrujillo16,PengLim16,RomanTrujillo17a,Amorisco18a,Amorisco18b,Chamba20}. 

One particular example found in a nearby galaxy group, NGC1052-DF2\footnote{This galaxy had been previously catalogued as [KKS2000]04, KKSG04, and PGC3097693 \citep{Karachentsev00}} stands out compared to any other known UDG (or even galaxy) in two key properties. First, its dark matter content (inferred using the kinematics of its GCs) is $\sim 400$ times lower than typical galaxies with the same stellar mass, $\Mstar \sim 10^8\Msun$ \citep{vanDokkum18a}. Second, the mass distribution of its GC population is very peculiar. It hosts 11 (spectroscopically confirmed) GCs with a narrow GC mass function (GCMF) that is shifted towards higher masses \citep[median mass $\sim 8\times10^5\Msun$;][]{vanDokkum18b} compared to the (nearly) universally observed behaviour across all galaxy types \citep{Rejkuba12}. The GCs constitute $\sim 5$ per cent of the total stellar mass of the galaxy, offsetting DF2 upwards from the mean observed relation between GC specific frequency and luminosity \citep{Peng08} by a factor of $\sim 3$ (see Table~\ref{tab:table1}). Due to its exceptionally low halo mass, the system also falls orders of magnitude off the relation between the GC system mass and the galaxy DM halo mass \citep{vanDokkum18b}. Recent spectroscopic data confirm that both the GCs and the field stars have ages $\sim 9\Gyr$, and find no evidence of variation in the metallicity across the GCs \citep{Fensch19}. 

The origin of GCs is still highly debated. This is in part due to the fact that they are usually quite old, making it extremely difficult to directly observe their formation conditions and environments at high redshift. However, much progress has been made in the last decade, with new models reproducing several of the characteristics of GC populations. In particular, many properties of globular clusters can be understood in a theoretical framework where they constitute the surviving evolved remnants of young massive star clusters formed in the early universe \protect\citep[e.g.,][]{KravtsovGnedin05, Elmegreen10, Shapiro10, Kruijssen15b, emosaicsI, Usher18, Choksi18}.  In this framework, the GCMF encodes key information about the high-redshift birth conditions (e.g., gas pressure or surface density, and disc angular speed) of the surviving clusters \citep{Reina-Campos17, Trujillo-Gomez19}. 
 
It may soon be possible to do galactic archaeology with GCs by using the current properties of GC systems (such as masses, ages, metallicities, and kinematics) to infer the  environmental conditions in their progenitor galaxies at the time of their formation. Recent works have used particle tagging or analytical cluster models \citep{Kruijssen15b, Reina-Campos17} and their implementation into fully cosmological galaxy formation simulations to make important steps towards this goal \citep{Renaud17, emosaicsI, emosaicsII, kraken, Ramos-Almendares20, Halbesma20, krakenII}. Semi-analytic GC formation models are now able to predict the GC mass, metallicity, and age distribution of large statistical samples of galaxies in DM-only cosmological simulations \citep{Choksi18, Choksi19a}.  In this paper we apply the analytical model for the initial cluster mass function from \citet{Trujillo-Gomez19} to investigate the properties of the progenitor of DF2 when its peculiar GC population formed.

Surprisingly, further searches for UDGs in the NGC1052 system recently led to the discovery of a galaxy very similar to DF2. NGC1052-DF4 also has a large size, an unusually massive GC population, and low dark matter content \citep{vanDokkum19}. The presence of both objects in the same galaxy group indicates that they might be part of a broader, undetected population of galaxies. 
 
It should be noted that the two peculiar properties of DF2, i.e., its low DM content and its atypically massive GC system, depend on the inferred distance of $D = 19 \pm 1.7\Mpc$ used by \citet{vanDokkum18a}. This value has recently been disputed by another group, who find instead $\sim 13 \Mpc$ \citep{Trujillo19, Monelli19}. Using the smaller distance, \citet{Trujillo19} conclude that DF2 is a dSph galaxy with a typical DM content and a normal GC luminosity function. However, the original distance estimate seems to withstand further scrutiny. In an independent analysis, \citet{BlakesleeCantiello18} confirmed it using the surface brightness fluctuations (SBF) technique. \citet{Cohen18} again obtained a similar distance using SBF, and \citet{vanDokkum18c} found further confirmation using a distance ladder anchored on a satellite of a megamaser host galaxy. In addition, the $\sim 20\Mpc$ distance to NGC1052-DF4 (the second galaxy hosting unusually massive GCs) was confirmed by \citet{Danieli20}\footnote{After submission of this work, \citet{Shen21b} obtained the most precise distance estimate so far, $D = 22.1\pm1.2\Mpc$, confirming the fiducial distance assumed in our analysis (within the uncertainties). }.  
  
In this paper we assume the original parameters for DF2 and its GC system as published by \citet{vanDokkum18a} and \citet{vanDokkum18b}, and confirmed independently by \citet{BlakesleeCantiello18}. We also evaluate in detail the effect of the uncertainties in the distance determinations on the predictions made in this work (see Appendix~\ref{sec:distance}). The paper is organised as follows. Section \ref{sec:Model} summarises the analytical model for the initial cluster mass function. Section \ref{sec:GCMF} presents the analysis of the observed GCMF to recover the ICMF. Section \ref{sec:conditions} describes the implications of the analysis for the star-forming conditions in DF2 at $z>1$. Section~\ref{sec:formation} discusses the formation scenarios that may have led to the recovered conditions, and Section \ref{sec:conclusions} presents a summary of our findings.

\section{Environmental dependence of the ICMF}
\label{sec:Model}

Here we present a brief description of the model for the initial cluster mass function (ICMF) and refer the reader to \citet[][hereon TRK19]{Trujillo-Gomez19} and \citet{Reina-Campos17} for further details. The ICMF is modelled as a power-law with an exponential truncation at the minimum and at the maximum cluster masses,
\begin{equation}
    \frac{{\rm d}N}{{\rm d}M} \propto M^{\beta} \exp\left( -\frac{\Mmin}{M} \right) \exp\left( -\frac{M}{\Mmax} \right) ,
    \label{eq:ICMF}
\end{equation}
where $M$ is the star cluster mass, $\beta = -2$ is the result of gravitational collapse in hierarchically structured molecular clouds \citep[e.g.][]{ElmegreenFalgarone96,Guszejnov18}, and $\Mmin$ and $\Mmax$ are the minimum and maximum truncation masses respectively. As shown by \citetalias{Trujillo-Gomez19}, this functional form provides an excellent representation of the observed mass functions of young clusters. Its variable minimum and maximum mass allow it be suitably adjusted to a wide range of galactic environments.

The maximum cluster mass $\Mmax$ is determined in the \citet{Reina-Campos17} model by finding the maximum mass within the shear-limited region of the disc \citep[the so-called ``Toomre mass'', see][]{Toomre64} that is able to collapse before star formation is halted by stellar feedback. In galactic environments with high gas surface densities or high angular rotation speeds, the centrifugal forces dominate and the Toomre criterion \citep{Toomre64} determines the maximum GMC mass and hence the maximum cluster mass. In environments with low gas surface density and low rotation speeds, stellar feedback stops the gas supply and reduces the fraction of the Toomre volume that can form stars. Under the assumption of a disc in hydrostatic equilibrium, the maximum cluster mass depends on the global environmental conditions as described by the mean ISM surface density, the disc angular rotation speed, and the Toomre $Q$ stability parameter. 

For the minimum cluster mass, $\Mmin$, we employ the model developed in \citetalias{Trujillo-Gomez19}. This model is based on the regulation of star formation in hierarchically structured molecular clouds by stellar feedback. It predicts the cloud mass range for which enough gas is converted into stars such that the structure can remain bound after the residual gas is expelled by feedback. Using the empirical result that clouds are substructured hierarchically \citep{Rosolowsky08,Heyer15,Henshaw20}, the bottom of the stellar structure merger hierarchy is defined by the largest cloud mass that can collapse into a single bound object. Populating a disc in hydrostatic equilibrium with spherical clouds results in a prediction for $\Mmin$ as a function of the galactic environment, i.e. the average ISM surface density $\SigmaISM$, angular rotation speed $\Omega$, and Toomre $Q$.

The predicted mass scales from these models have been found to reproduce the observed cluster mass distributions in a variety of galactic environments in the local Universe. \citet{Reina-Campos17} showed that the predicted $\Mmax$ agrees very well with the largest observed cluster masses in a broad range of environments (and as a function of galactocentric radius) including the conditions in the Milky Way, M31, and a typical $z=2$ clumpy star-forming galaxy. \citetalias{Trujillo-Gomez19} further compared the predictions for the full ICMF with observations in the solar neighbourhood, the Central Molecular Zone of the Milky Way, the LMC, M31, the Antennae interacting galaxies, and the starbursting core of M82. The predictions for the low-mass truncation agree well with the upper limits on the minimum cluster mass inferred from observations of these galactic environments. At the high gas surface densities typical of GC formation conditions at high redshift, $\SigmaISM \ga 10^2\Msunpc2$, the minimum cluster mass depends only on the mean ISM surface density, with a scaling $\Mmin \propto \SigmaISM^3$. At even larger gas surface densities, $\SigmaISM > 2\times10^3\Msunpc2$, the scaling becomes increasingly steeper, and above $4\times10^3\Msunpc2$ the entire cloud mass hierarchy collapses into a single bound cluster with mass limited by $\Mmax$. In the regime where high-redshift galaxies are found, $\SigmaISM \ga 10^2\Msunpc2$, feedback becomes inefficient at preventing the collapse of the most massive molecular clouds, and the maximum cloud mass corresponds to the largest shear-unstable region, given by the Toomre mass, $\MToomre = 4\pi^5 G^2\SigmaISM^3/\kappa^4$, where $\kappa$ is the epicyclic frequency of the disc. The maximum mass of a star cluster is then obtained from the product of the cluster formation efficiency (CFE) $\Gamma$, the integrated star formation efficiency $\epsilon$, and the maximum cloud mass, 
\begin{equation}
    \Mmax = \Gamma(\SigmaISM,\kappa,Q) \times ~\epsilon~ \times 4\pi^5 G^2\SigmaISM^3/\kappa^4 , 
    \label{eq:Mmax}
\end{equation}
where $\Gamma$ is the CFE obtained using the \citet{Kruijssen12d} model, and $\epsilon = 0.1$ \citep[e.g.][]{Lada03,Oklopcic17}. 

The dominant uncertainties in the ICMF model lie in the assumed values of the integrated cloud-scale star formation efficiency $\epsilon$, the star formation efficiency per free-fall time $\epsff$, the gas conversion efficiency in pre-stellar cores $\epscore$, and the time-scale for the onset of feedback $\tsnO$. The values assumed here, $\epsilon=0.1$, $\epsff=0.01$, $\epscore=0.5$, and $\tsnO=3\Myr$, are broadly consistent with a variety of observations (for a detailed discussion of the ICMF uncertainties see \citealt{Reina-Campos17} and \citetalias{Trujillo-Gomez19}). Assuming a factor of 2 larger $\epsff$ (or a factor of 2 smaller $\epscore$) increases the predicted minimum cluster mass for $\SigmaISM = 10^2\Msun$ by a factor of $\sim 7$ in the model. At this gas surface density, assuming $\tsnO = 1\Myr$ (instead of the fiducial $3\Myr$) reduces the minimum mass by $\sim 50$ per cent. The sensitivity of $\Mmin$ to these parameters grows at larger gas densities. In the same regime, doubling the assumed value of $\epsilon$ doubles the predicted maximum cluster mass. Although the quantitative predictions presented here are sensitive to these parameter choices, the qualitative dependence of the minimum and maximum masses on the gas pressure and shear remain unchanged.

A particularly interesting prediction of the ICMF model is that across the range of conditions spanned by observed galaxies, there should be a strong variation (of several orders of magnitude) in the minimum and maximum cluster mass, as well as in the resulting total width of the ICMF. This environmental dependence is driven mainly by the ISM surface density $\SigmaISM$ and the angular speed $\Omega$, and results in very different predictions for the ICMF across the parameter space spanned by observed galaxies (see figure 5 in \citetalias{Trujillo-Gomez19}). In particular, environments with large gas surface density and strong centrifugal support produce very narrow ICMFs with high median cluster masses. These conditions are typical of starbursting nuclear rings and the Central Molecular Zone of the Milky Way, where the model predicts a very narrow ICMF with $\Mmin \sim 3.2\times10^3\Msun$ and a width of less than 1 dex \citepalias{Trujillo-Gomez19}.

The \citetalias{Trujillo-Gomez19} model fully determines the shape of the initial cluster mass distribution, but its normalization is set by the bound fraction of star formation, which can be obtained using the \citet{Kruijssen12d} model. In this model, the fraction of star formation in  gravitationally bound clusters is determined by the hierarchical nature of the ISM density structure, where the highest density regions have the shortest free-fall times and can convert enough gas into stars to remain bound after gas expulsion by feedback. The \citet{Kruijssen12d} cluster formation efficiency includes early cluster disruption due to the `cruel cradle' effect. As we are interested in the birth cluster conditions, we neglect this effect and use `CFE' and `bound fraction' interchangeably to refer to the naturally bound fraction of star formation. Using the bound fraction before accounting for cluster dynamical mass loss is a conservative assumption and provides a strict lower limit for the initial mass under the ICMF. Assuming a disk in hydrostatic equilibrium, the ISM density PDF is set by the galactic environment, such that the CFE can be predicted from the same global properties used for the ICMF, $\SigmaISM$, $\Omega$, and Toomre $Q$.

\section{The ICMF of the observed globular clusters in DF2}
\label{sec:GCMF}

We now proceed to apply the \citetalias{Trujillo-Gomez19} ICMF model and the \citet{Kruijssen12d} CFE model to the GCMF of DF2 to reconstruct the star formation environment that produced its peculiar cluster population $\ga 9\Gyr$ ago. The procedure can be summarised as follows:

\begin{enumerate}

\item Assuming that the 11 GCs formed in a single burst of star formation \citep[due to the narrow spread in their iron abundances;][]{Fensch19}, obtain the GCMF at the time of their formation (the birth GCMF) by correcting the observed GCMF for the fraction of mass lost by stellar evolution.

\item Fit the birth GCMF with the model ICMF (equation~\ref{eq:ICMF}) using the Maximum Likelihood Estimator (MLE) to obtain its two parameters, the minimum and maximum truncation masses $\Mmin$ and $\Mmax$, as well as its normalisation. Since lower mass clusters may have been disrupted after birth, the value of $\Mmin$ becomes the upper limit on the minimum cluster mass of the initial cluster population of DF2.
    
\item Obtain the lower limit on $\Mmin$ using the constraint that the integrated stellar mass under the ICMF cannot exceed the total present-day stellar mass of DF2\footnote{We assume that the fraction of stellar mass in GCs has not changed significantly in DF2 (after its GCs were formed) due to tidal stripping by NGC1052. This would be the case if GCs and stars have a similar spatial distribution, and seems to be supported by the rough similarity between the effective radii of the GC system and the galaxy (see Table~\ref{tab:table1}). Furthermore, \citet{vanDokkum18a} find no evidence of morphological asymmetry or tidal features, suggesting that the galaxy has survived in its present form for longer than a few dynamical times.} (corrected for stellar evolution). Since the galaxy likely had an extended star formation history, the minimum mass will always be larger that this strict lower limit.

\item While assuming the fixed $\Mmax$ value obtained above, search the $(\SigmaISM,\Omega)$ parameter space for the value of $\Mmin$ that yields a ratio of stellar mass in clusters (the integrated ICMF mass) to the total mass of the galaxy (i.e., the CFE) which matches the prediction by \citet{Kruijssen12d} for the same combination of ($\SigmaISM$, $\Omega$). Since there is no constraint on Toomre $Q$, find the solution for a representative range of values of this parameter.  
    
\item The minimum and maximum masses have different environmental dependence. Therefore, a physical solution corresponds to the small region where the best-fitting values overlap in the $(\SigmaISM,\Omega)$ space. For a choice of Toomre $Q$, this region of parameter space determines the mean values and uncertainties of $\SigmaISM$ and $\Omega$.

\end{enumerate}

Table~\ref{tab:table1} summarizes the observed and inferred properties of NGC1052-DF2 adopted here.
\begin{table}
    \centering
	\begin{tabular}{lc} 
		\hline 
		 Property        & Value \\
		\hline
		 $L_V$           & $1.1\times10^8\Lsun$ \\
         $\Mstar$        & $2\times10^8\Msun$ \\
         $r_e$           & $2.2\kpc$  \\
         $r_e^{\rm GCs}$ & $3.1\kpc$ \\
         $N_{\rm GC}$    & 11 \\
         $S_N$           & 8.5  \\
		\hline 
	\end{tabular}
	\caption{Properties of NGC1052-DF2 adopted in this work, assuming a distance of $20\Mpc$. The values are taken from \citet{vanDokkum18a,vanDokkum18b}, and the specific frequency is defined by the equation $S_N = 8.51\times10^7~N_{\rm GC}/(L_V/L_{V\odot})$ \citep{Harris13}.}
	\label{tab:table1}
\end{table}
First, to obtain the ICMF of the clusters that evolved into the GC population of DF2, cluster evolution must be taken into account. To estimate the birth masses of the GCs we use the observed $V$-band luminosities with a mass-to-light ratio $M/L_V \approx 2$ \citep{vanDokkum18b}, and correct for the fractional mass loss due to stellar evolution assuming a \citet{Kroupa01} IMF, $f = 0.35$ \citep{Lamers10}. To keep our estimates conservative, we do not correct for dynamical mass loss in the observed clusters. Dynamical mass loss corrections are highly uncertain and would have the systematic effect of shifting the low-mass truncation of the ICMF towards larger values. However, the relative dynamical mass loss decreases towards higher GC masses. For the GC masses $>10^{5.8}\Msun$ in DF2, models predict less than 30~per~cent of the initial mass is lost through dynamical processes \citep[e.g.,][]{Kruijssen15b, Reina-Campos18}. The GC birth masses are $[1.4,  0.96, 1.0, 2.9, 1.8, 1.3, 1.3, 0.72, 1.5, 0.79,  0.72] \times10^6\Msun$, with a median mass $\tilde{M} = 1.3\times10^6\Msun$. Since the ages of the GCs and the field stars are very similar \citep{Fensch19}, we apply the same stellar evolution correction to the total stellar mass, and obtain the birth mass of the field stars, $\Mstar^{birth} = 3.1\times10^8\Msun$.

Following the analysis of the Fornax dSph in \citetalias{Trujillo-Gomez19}, the uncertainty in $\Mmin$ due to the disruption of low-mass clusters is bracketed between two limits resulting from the unknown fraction of stars in the star formation burst that formed in bound clusters (i.e., the CFE). First, in the {\it minimum CFE} model only the stars currently observed in clusters were born clustered, implying that the GCs observed today are the only clusters that formed and that the ICMF is simply the GCMF corrected for stellar evolution. This gives an upper limit to $\Mmin$. Second, in the {\it maximum CFE} model all the stars in the galaxy were born in clusters, implying that the integrated ICMF should at most equal the total coeval stellar mass of the galaxy. This yields a lower limit on $\Mmin$. Since there are no constraints on the star formation history of DF2, we follow the assumption used by \citet{vanDokkum18b} based on the similarity of their average colours, that the field stars have similar ages to the GCs. A recent spectroscopic analysis confirms this assumption \citep{Fensch19}.

To constrain $\Mmin$ in each of the two models we first fit the birth GCMF using equation~(\ref{eq:ICMF}). This is done using the Maximum Likelihood Estimator \citep{Fisher1912} with uniform priors on $\Mmin$ and $\Mmax$ in the allowed region where $\Mmax \geq \Mmin$ and the total mass under the ICMF matches the total mass of the 11 GCs within the observational errors quoted by \citet{vanDokkum18b}. The best-fit values of $\Mmin$ and $\Mmax$ become the parameters of the ICMF in the {\it minimum CFE} model. In the {\it maximum CFE} model, we reduce $\Mmin$ while keeping $\Mmax$ fixed (at the best-fit value of the birth GCMF) until the total mass under the ICMF matches the total stellar mass of the galaxy. Our model does not explicitly include ICMF sampling uncertainties, but the MLE method automatically accounts for this effect by selecting the most likely ICMF given the observed distribution of cluster masses. 

\begin{figure}
    \includegraphics[width=\columnwidth]{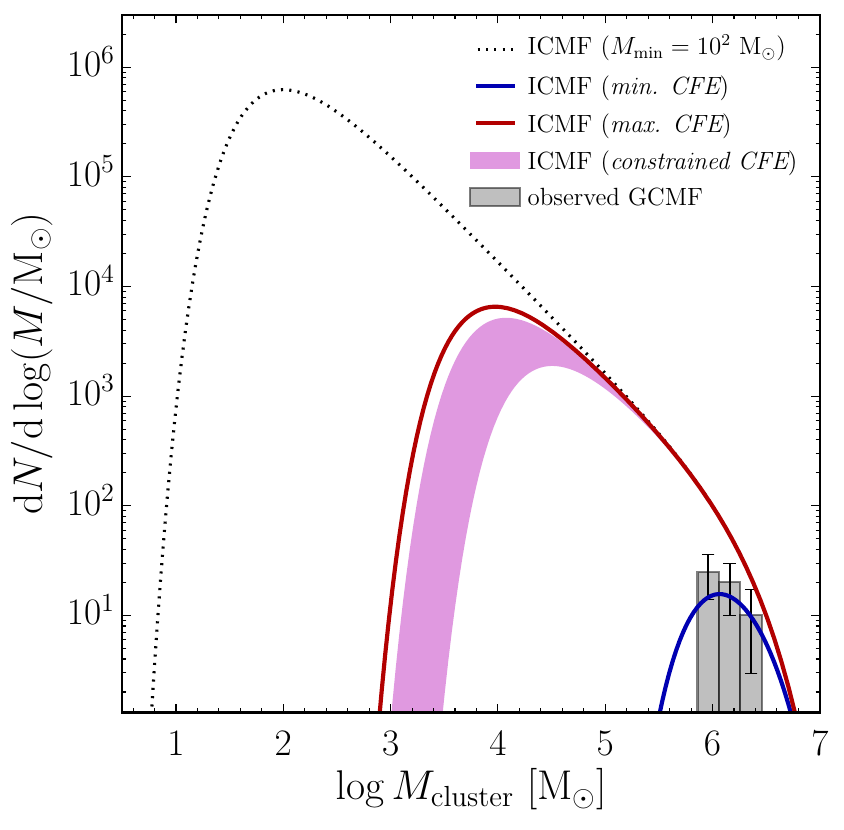}
    \caption{Constraints on the ICMF of the observed GCs in DF2. The shaded histogram is based on the 11 GCs identified by \citet{vanDokkum18b} and was corrected for the effects of stellar evolution \citep{Lamers10} to obtain the {\it birth} masses. The model curves bracket the allowed range of the ICMF in the galaxy at the epoch of GC formation $\ga 9 \Gyr$ ago. The blue curve shows the {\it minimum CFE} model, where the GCs observed today were the only ones that formed. The red curve is the ICMF using the {\it maximum CFE} model where all the field stars in the galaxy formed within clusters which later dissolved. The {\it constrained CFE} model (shaded band) was obtained by matching the CFE predicted by the \citet{Kruijssen12d} model based on the gas surface density and angular velocity for $0.5 \leq Q \leq 3.0$. The dotted line shows the ICMF for a minimum cluster mass of $10^2\Msun$, which is commonly assumed in studies of young cluster populations in nearby galaxies. This is ruled out, because DF2 does not contain enough field stars.}
\label{fig:CMF_DF2}
\end{figure}

The resulting ICMF models are shown in Figure \ref{fig:CMF_DF2}. For the {\it minimum CFE} model we  obtain $\Mmin = 1.9^{+0.6}_{-0.5}\times10^6\Msun$, and $\Mmax = 1.9^{+1.5}_{-0.5}\times10^6\Msun$, where the uncertainties correspond to the values where the likelihood function decreases by a factor of $1/e$. The close match between the minimum and maximum truncation masses simply indicates that the observed DF2 GCMF is very narrow. For the {\it maximum CFE} model, the minimum mass is reduced to $\Mmin = 9.5^{+2.8}_{-2.6}\times10^3\Msun$. Lastly, for the self-consistent {\it constrained CFE} model we assumed a range of Toomre $Q$ values $0.5 \leq Q \leq 3.0$ (shaded band in Figure~\ref{fig:CMF_DF2}), and obtain solutions in the range $\Mmin = 1.2{-}2.1 \times10^4\Msun$ for the preferred ($Q=0.5$) model. This is the typical value of Toomre $Q$ in galaxies with GC formation conditions at $z \sim 2$, which display values in the range $Q \approx 0.2{-}1.6$  outside their nuclear regions \citep{Genzel14}. The cluster formation efficiency for this model is $[87,78]$ per~cent for $Q$ in the range $[0.5,3.0]$, implying that the ISM conditions required to produce the observed GCMF also result in a very large fraction of the total stellar mass of DF2 forming in bound clusters. Finally, we note that the total stellar mass of DF2 firmly rules out the minimum cluster mass that is typically assumed in nearby star-forming disc galaxies of $10^2\Msun$ \citep[e.g.,][]{Lada03,Adamo20}.

Evidently, there are many caveats involved in reconstructing the ICMF from the observed GC masses. Various dynamical processes including two-body relaxation and tidal shocking by molecular clouds are expected to dissolve or reduce the masses of clusters over many billions of years \citep[e.g.,][]{OstrikerGnedin97, BaumgardtMakino03, Kruijssen15b}. These mechanisms might still be important in DF2 but are not relevant to our results because the star-forming conditions that determine the minimum cluster mass and the CFE are obtained under the most conservative assumption of negligible dynamical mass loss. The predictions for the environmental conditions are then lower bounds on the true values of $\SigmaISM$ and $\Omega$. Section~\ref{sec:conditions} discusses the effect of assuming non-negligible mass loss of the surviving GCs. Moreover, neither DF2 nor DF4 appear to contain a nuclear star cluster, which indicates that dynamical friction has not affected the most massive clusters.

\section{The galactic environment that produced the massive GCs in DF2}
\label{sec:conditions}

\begin{figure*}
    \includegraphics[width=0.90\textwidth]{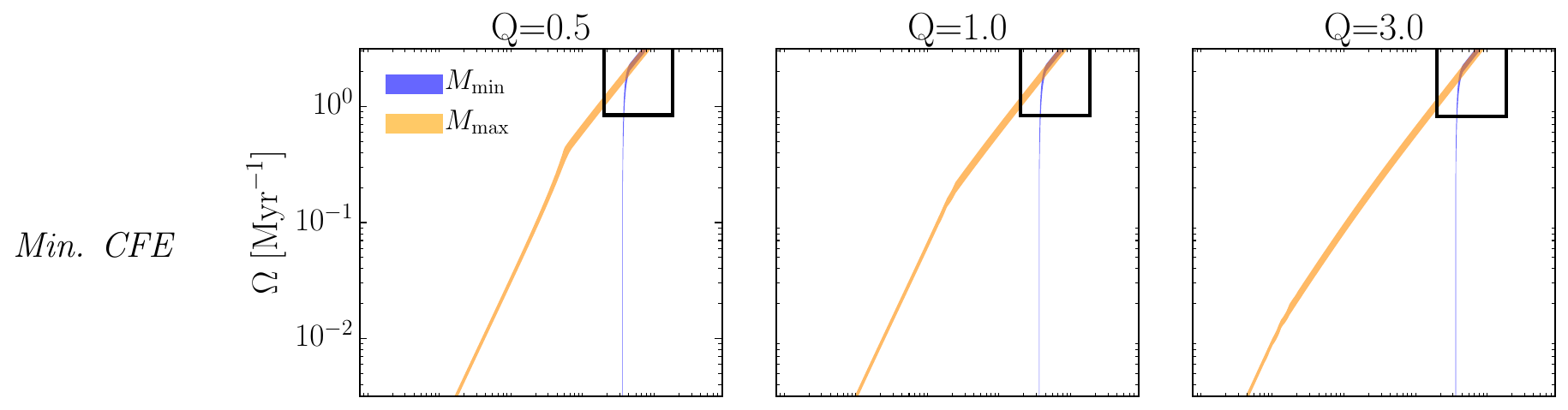}
    \includegraphics[width=0.90\textwidth]{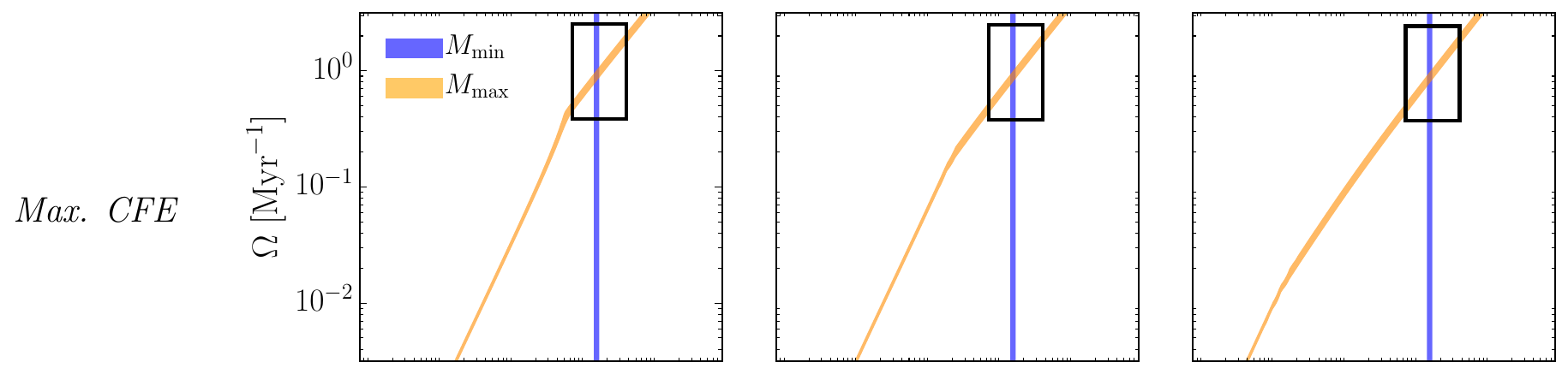}
    \includegraphics[width=0.90\textwidth]{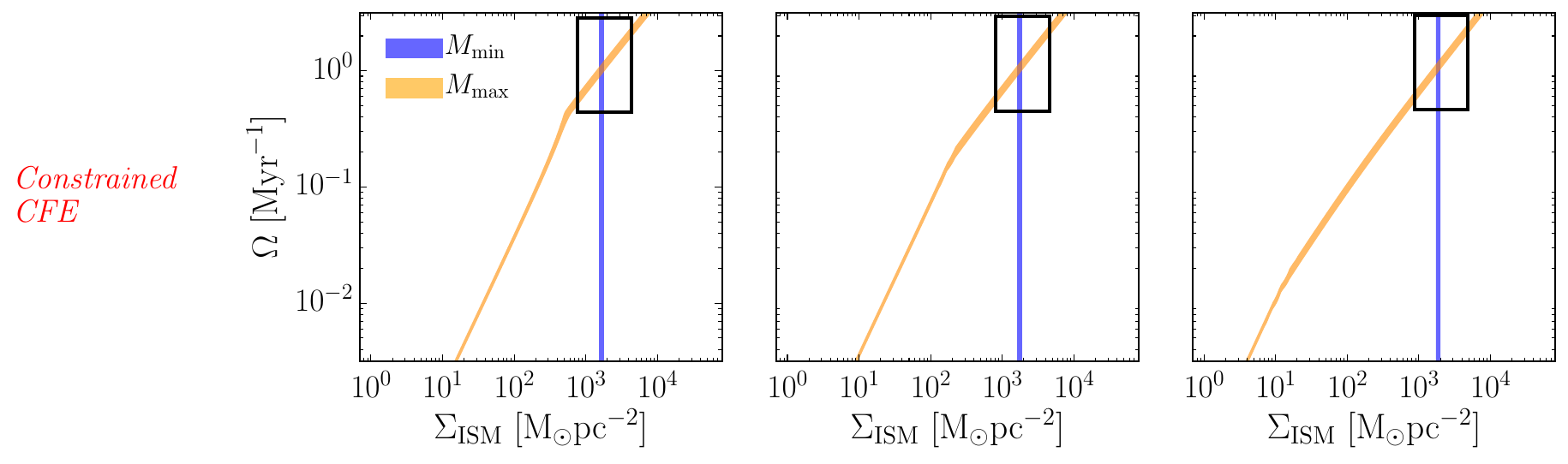}
    \caption{Birth galactic environmental conditions required by the observed GCMF in the ultra-diffuse galaxy NGC1052-DF2. The shaded regions show the locations in the parameter space of gas surface density and disc angular velocity that give rise to the observed minimum (blue) and maximum (orange) truncation masses of the ICMF shown in Figure \ref{fig:CMF_DF2}. The top and middle rows show the two bracketing models corresponding to the {\it minimum} and {\it maximum} possible CFE values, respectively. The bottom row shows the model with the CFE constrained by the \citet{Kruijssen12d} model. From left to right, the columns show the results for values of Toomre $Q \in [0.5,1,3]$ respectively. Since the minimum and maximum truncation masses in the ICMF are independently predicted by the \citetalias{Trujillo-Gomez19} model, physical solutions correspond exclusively to the regions where their conditions overlap (indicated by the boxes). This indicates that the progenitor of DF2 had very high gas surface density and shear when its GCs formed, $\ga 9\Gyr$ ago.}
\label{fig:DF2_conditions}
\end{figure*}

The \citetalias{Trujillo-Gomez19} model predicts a strong variation of the minimum and maximum truncation mass of the ICMF due to the influence of galactic environment. Having derived the ICMF of the DF2 progenitor, we proceed to find the global star-forming conditions at the time of GC formation, that led to such a narrow ICMF with a high mean mass. Figure~\ref{fig:DF2_conditions} shows the mean ISM surface density and angular velocity of the galaxy that produced the {\it constrained CFE} ICMF in Figure \ref{fig:CMF_DF2}. The figure also shows the conditions that give rise to the {\it minimum} and {\it maximum CFE} bracketing cases for three representative values of Toomre $Q$ in the range $0.5 \leq Q \leq 3.0$. These represent the full range of conditions that are consistent with CFE values $5{-}100$ per~cent as required by the total stellar mass of the galaxy. Due to the different dependence of the minimum and maximum masses on the environmental conditions \citepalias[see][]{Trujillo-Gomez19}, only regions where the solutions for $\Mmin$ and $\Mmax$ overlap are physically meaningful.

Figure~\ref{fig:DF2_conditions} indicates that, regardless of the uncertainty in the CFE and in the Toomre $Q$ parameter, the gas environment in this ultra-diffuse galaxy must have been quite extreme, i.e. high gas surface density $\SigmaISM \ga 10^3\Msunpc2$ and shear $\Omega \ga 0.7\Myr^{-1}$, to form the observed population of massive GCs. In the self-consistent {\it constrained CFE} model, the mean values are $\SigmaISM = 1.7\times10^3\Msunpc2$ and $\Omega = 1.0\Myr^{-1}$ for the preferred ($Q=0.5$) model. In this model, 87 per cent of all stars are formed in bound clusters (Section~\ref{sec:GCMF}).  

The environmental conditions recovered by the model are extreme and lie orders of magnitude above the values found in typical galaxies in the local universe, where $\SigmaISM \lesssim 50\Msunpc2$ and $\Omega \lesssim 0.05\Myr^{-1}$ \citep[e.g.,][]{Leroy08}. At $z\sim0$ they are more commonly found in circumnuclear starbursts and mergers \citepalias[see][]{Trujillo-Gomez19}, and match closely the environment found in the central $\sim 100\pc$ of the Milky Way \citep{Kruijssen15a, Henshaw16}. The extreme conditions that were required to form the observed GC population contrast strongly with the current diffuse nature of the host galaxy, and might indicate that DF2 formed its GCs either in a nuclear starburst, or during a gas-rich major merger event in its early history. In the starburst scenario, the high gas surface density $\SigmaISM \sim 1.7\times10^3\Msunpc2$, and shear $\Omega \sim 1.0\Myr^{-1}$, immediately point to a very compact galactic progenitor.   

The analysis presented here relies on the most conservative assumption of negligible dynamical mass loss of the surviving GCs. Accounting for dynamical mass loss to recover the birth GCMF would increase the lower bound on the CFE above the value of $\sim5$ per~cent that is used in this analysis. As a result, the minimum cluster mass of the {\it maximum CFE} model would increase (to avoid overproducing the galaxy stellar mass), leading to a shift in the solution towards more extreme ISM conditions than those obtained in our analysis (i.e. higher gas surface densities and angular speeds in Figure~\ref{fig:DF2_conditions}).

\section{Implications for the formation of DF2}
\label{sec:formation}

The unique location of the environmental conditions for the formation of the GCs in DF2 in the surface density and angular velocity plane require its $z \ga 1.3$\footnote{This is the redshift (assuming the \citealt{Planck18} cosmological parameters) corresponding to the lower limit on the GC ages of $\sim 9\Gyr$ from \citet{vanDokkum18b}.} progenitor to contain a region of high gas pressure and shear large enough to form $78-87$ per cent of its stars in bound clusters (see Section~\ref{sec:conditions}). As discussed in Section~\ref{sec:conditions}, these could arise in two ways: as a result of a nuclear starburst, or a gas-rich major merger. In this section we examine the implications of each of these scenarios. In addition, we consider the possible role of feedback-driven gas compression. From here on we assume the ICMF and galactic environmental conditions corresponding to the {\it constrained CFE} model with $Q=0.5$.

\subsection{Formation scenarios}

\subsubsection{Scenario I: nuclear starburst}
\label{sec:starburst}

Assuming that the ISM conditions derived above occur in a rotating disc in hydrostatic equilibrium, they are comparable to those in the nuclei of starbursting galaxies in the local Universe. These conditions occur within a normal disc galaxy if internal torques drive gas towards the centre, producing a large concentration of baryons (gas and/or stars), which in turn increases the angular frequency and shear in the nuclear region. Since the highest global shear values occur near the galaxy centre, this scenario directly translates into a constraint on the enclosed mass in the centre of DF2 during the formation of its massive GCs, $\ga 9\Gyr$ ago. 

The angular speed, $\Omega(r)$ is a proxy for average enclosed density,
\begin{equation}
    \Omega(r) = \sqrtsign{\frac{G M_{\rm tot}(<r)}{r^3}} ,
\label{eq:omega}
\end{equation} 
while the average mass surface density is given by
\begin{equation} 
    \overline{\Sigma}_{\rm tot} = \frac{ M_{\rm tot}(<r) }{ \pi r^2 }  .
\label{eq:sigma}
\end{equation} 
Assuming that the gas is the dominant mass component, $M_{\rm tot} \simeq M_{\rm gas}$, these two equations can be solved simultaneously using the values of gas surface density and angular frequency obtained in Section~\ref{sec:conditions} to obtain the density and size of the starburst region where DF2's GCs formed. Including the DM contribution to the mass in equations (\ref{eq:omega}) and (\ref{eq:sigma}) requires an extra constraint on $M_{\rm DM}(<r)$. This is provided by the NFW \citep{navarro97} parametrization of the DM density profile as a function of halo mass and concentration, 
\begin{equation}
    \rho_{\rm NFW}(r) = \frac{4\rhos}{r/\rs \left(1 + r/\rs \right)^2} ,
    \label{eq:NFW}
\end{equation}
where 
\begin{equation}
    \rs \equiv \Rvir/c
\end{equation}
is the scale radius, $\rhos$ is the density at the scale radius, and $c$ is the halo concentration. The virial radius is
\begin{equation}
    \Rvir \equiv \frac{3\Mhalo}{4\pi (200\rhocrit)} ,
    \label{eq:rvir}
\end{equation}
where $\rhocrit$ is the critical density of the Universe. The density at the scale radius and the concentration are related through the equation
\begin{equation}
    \rhos = \frac{200\rhocrit}{12} \frac{c^3}{\left[ \ln(1+c) - c/(1+c)\right]} .
\end{equation}
Assuming the upper limit on the virial mass of DF2 from \citet{vanDokkum18a}, $\Mhalo \la 1.5\times10^8\Msun$, and the average concentration-mass relation from \citet{DuttonMaccio14},
\begin{equation}
    \log c(\Mhalo) = 0.905 - 0.101\log \left(\frac{\Mhalo}{10^{12}\Msunh}\right) ,
    \label{eq:cm_relation}
\end{equation}
we can obtain the DM density profile. This estimate assumes that the halo mass has not evolved in the last $\sim 9 \Gyr$, but as we demonstrate below, the results are insensitive to this assumption.

\begin{figure}
    \includegraphics[width=\columnwidth]{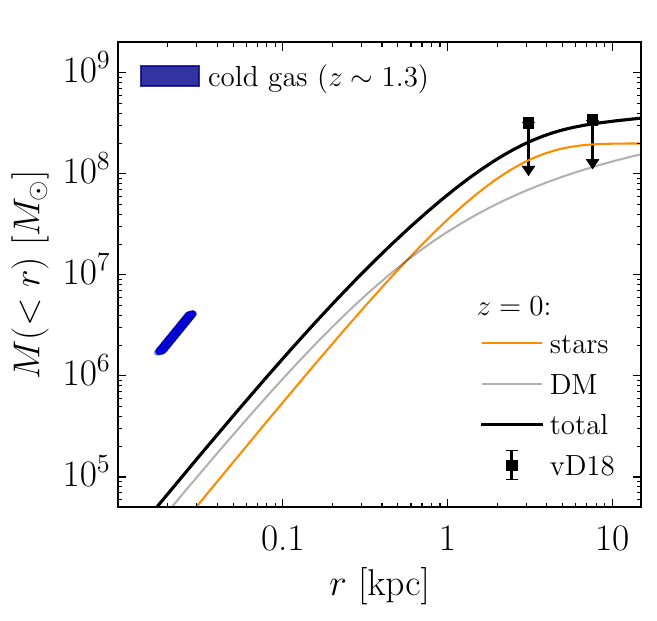}
    \caption{Present-day density profile of DF2 compared to the galactic environment required in the starburst scenario (Section~\ref{sec:starburst}) at $z\sim1.3$. The blue shaded area indicates the parameter region of enclosed mass and galactocentric radius that represents solutions to equations (\ref{eq:omega}) and (\ref{eq:sigma}) for the gas surface density and angular frequency in the {\it constrained CFE} model assuming $Q=0.5$. Constraints on the present-day mass distribution in the galaxy from \citet{vanDokkum18a} are also shown for comparison. The grey line corresponds to the DM, while the orange and black lines show the stars and the total mass. The formation of the massive GCs hosted by DF2 requires a very compact nuclear region of size $r\la30\pc$ dominated by very dense gas. }
\label{fig:DF2_density}
\end{figure}

The resulting constraints on the enclosed density and radial extent of the progenitor obtained using the model are shown in  Figure~\ref{fig:DF2_density}. The shaded region in the $r$ and $M(<r)$ parameter space represents the $1\sigma$ uncertainty range of solutions to equations (\ref{eq:omega}) and (\ref{eq:sigma}) for the $\SigmaISM$ and $\Omega$ region in the model. Shear values $0.9{-}1.2\Myr^{-1}$ and gas surface densities $1.5{-}1.8\times10^3\Msunpc2$ can be produced by a dense central gas nucleus of radius in the range $17{-}29\pc$ containing a total gas mass in the range\footnote{The fraction of enclosed DM mass would be very low, $\sim 2$ per cent (assuming $M_{\rm halo} = 1.5\times10^8\Msun$ for an NFW density profile of average concentration at $z=0$). Assuming a much larger initial halo mass, $\Mhalo \sim 10^{11}\Msun$ (the $z=0$ abundance matching estimate given the stellar mass of DF2) only changes the radial range by $\sim 6$ per cent because the central DM density does not increase significantly.} $1.7{-}4.1\times10^6\Msun$. Note that this gas mass would not be enough to produce even the total birth mass of the observed GCs (let alone the stellar mass of DF2, $\Mstar^{birth} = 3.1\times10^8\Msun$). To satisfy this constraint for the mean conditions requires that the $\sim 23\pc$ nuclear region continuously feed from an extended gas reservoir with an outer radius at least 10 times larger, $r > 230\pc$ (assuming a constant gas surface density profile). However, to avoid forming clusters with a higher ICMF truncation mass (due to the decrease in $\Omega$ across the extended region), the gas infall must be very rapid. Another remaining but very unlikely possibility is that each of the 11 GCs formed in a different progenitor galaxy with very similar ISM conditions. This makes a nuclear starburst a less favourable explanation for the GCMF of DF2. 

The environmental conditions inferred here are, however, surprisingly similar to those of the `Little Blue Dots' (LBDs) discovered by \citet{Elmegreen17}. These are extremely compact ($r \la 300\pc$), low-mass ($\Mstar < 10^{7.5}\Msun$) galaxies at $z>0.5$, which appear blue due to their extremely high specific star formation rates, up to $\sim 10^{-7}\Gyr^{-1}$. \citet{Elmegreen17} estimate that their gas mass could be up to $50$ times larger than their stellar mass, leading to gas pressures high enough to form the low-metallicity GCs hosted by many dwarf galaxies at present. The high star formation rates and gas densities of the most extreme LBDs at $z=1{-}2$ could account for the formation of DF2's entire massive GC population on a timescale $< 10\Myr$. 

Assuming that the gas supply was enough to form each of the GCs in a separate burst, the compact arrangement of the birth region of radius $r \la 30\pc$ implies that the GC orbits would evolve quickly due to dynamical friction. The time required for the GCs to spiral into the centre of the galaxy due to the effect of dynamical friction can be estimated using equations 8.2 and 8.12 of \citet{BinneyTremaine},
\begin{equation}
    t_{\rm fric} = \frac{19\Gyr}{\ln\Lambda}\left(\frac{r_{\rm init}}{5\kpc}\right)^2 \frac{\sigma}{200\kms}\frac{10^8\Msun}{M_{\rm GC}} ,
\end{equation}
where the Coulomb logarithm is
\begin{equation}
    \Lambda = \ln\left(\frac{b_{\rm max}}{\max(r_{\rm GC},GM/v_{\rm typ}^2)}\right) ,
\end{equation}
 $r_{\rm init} = 30\pc$ is the initial galactocentric radius of the GCs, $\sigma = \Vc(r_e)/\sqrt{2}$ is the velocity dispersion of the galaxy at the half-mass radius, $M_{\rm GC}=2.9\times10^6\Msun$ and $r_{\rm GC}\approx 6\pc$ are the birth mass and half-mass radius of the most massive GC, respectively, $b_{\rm max} \sim r_{\rm init}$ is the maximum impact parameter, $r_e = 2.2\kpc$ is approximately the half-mass radius of the galaxy, and $v_{\rm typ} \approx \Vc(r_e)$ is the typical velocity of the stars. For the total density profile at $z=0$ (shown as a black line in Figure~\ref{fig:DF2_density}), the total mass enclosed within the initial radius of formation of the GCs is lower that the mass of a single GC, i.e. $r_{\rm init}$ is well below the 90 degree deflection impact parameter $b_{90} \approx GM_{\rm GC}/v_{\rm typ}^2 = 41\pc$. The dynamical friction timescale is then $t_{\rm df} \rightarrow 0$, and if the most massive GC formed in such a compact region it would instantly become the centre of the galaxy. To prevent the formation of a nuclear star cluster by the coalescence of the GCs, their stellar feedback would have to rapidly blow out the remaining gas, reducing the galaxy potential and expanding the GC orbits. This is very unlikely given that feedback timescales in nearby galaxies are in the range $1{-}5\Myr$ \citep{Kruijssen19, Chevance20}, much longer than the coalescence time of the GCs. 

To summarise, assuming that the GCs formed in a nuclear starburst within a disc in hydrostatic equilibrium  leads to the requirement of multiple bursts fed by a continuous supply of gas to a very small and dense central region. However, this region must be so small that the GCs would rapidly become the centre of the galaxy and thus form a nuclear star cluster. The fine-tuning required in this scenario in addition to the absence of a nuclear star cluster in DF2 itself may then point to a different origin of its massive GCs.

\subsubsection{Scenario II: major merger}
\label{sec:merger}

As discussed above, the large minimum ICMF truncation mass requires very large gas surface densities (a proxy for gas pressure), while the fact that the maximum mass is not much larger than the minimum requires high galactic shear (a proxy for gas motions which stabilise larger regions against gravitational collapse). Galactic dynamics in an equilibrium disc is not the only way for these conditions to emerge in galaxies. They also commonly occur in major merger environments, driven by the orbital dynamics. Major mergers of gas-rich galaxies are known to be prolific sites for massive cluster formation in the local Universe \citep[e.g.,][]{Holtzman92,WhitmoreSchweizer95,Holtzman96,Zepf99}. Following the first pericenter passage, cold gas loses angular momentum through excitation of non-axisymmetric modes in the discs \citep{mihoshernquist96}. Combined with shocking from ram pressure and tides, this causes gas to be compressed to very high densities $\ga 10^3{-}10^4 \Msunpc2$ \citep[e.g.,][]{Karl10,Kruijssen12c,Renaud14,Kim18,Moreno19} in the filaments and galaxy nuclei before the final coalescence of the two galaxies. The conversion of orbital energy into turbulence increases the velocity dispersion in these filaments and clumps. Together, the dispersion and the shear from the large-scale motions may support the gas against gravitational collapse and reduce the maximum mass of a gravitationally unstable region (the equivalent of the Toomre mass in an equilibrium disc). 

In this scenario, the hydrostatic equilibrium assumption needed for the calculation of the maximum cluster mass in the  \citetalias{Trujillo-Gomez19} model is no longer valid. However, high gas pressures are still required by the large minimum mass because the underlying physics (runaway hierarchical merging of protoclusters) does not rely on the assumption of hydrostatic equilibrium. In mergers, the required gas pressures typically occur where the gas density is high \citep{Kruijssen12c, Lahen20}. As was shown in Section~\ref{sec:conditions} and Figure~\ref{fig:DF2_density}, at the high gas surface densities required by the minimum mass, the gas is gravitationally dominant. The maximum mass in the model is simply the fraction of the Toomre mass that can collapse before stellar feedback halts the process. At high pressures, feedback is inefficient at halting the collapse of the unstable region and the maximum mass is determined only by the product of the cloud-scale star formation efficiency $\epsilon$, the CFE $\Gamma$, and the Toomre mass $\MToomre$ (see Eq.~\ref{eq:Mmax}). The epicyclic frequency $\kappa$ now traces local shearing motions during the merger, and can be calculated for an arbitrary mass configuration using the eigenvalues of the local tidal tensor without the need to assume hydrostatic equilibrium \citep{emosaicsI}. This implies that the qualitative predictions of the \citetalias{Trujillo-Gomez19} model could still be applied in merger environments, indicating that a major merger may have produced the DF2's massive GCs.  

The formation of clusters in dwarf galaxy mergers was studied recently by \citet{Lahen19a,Lahen20} using isolated disc simulations with sub-parsec resolution. For a fixed gas surface density, they report CFE values even larger than those predicted by the \citet{Kruijssen12d} model used here. In these simulations, the merger induces pressures large enough to form $\sim 10^6\Msun$ bound clusters in gas-rich dwarf galaxies with stellar masses $\sim 10^7\Msun$. In simulations that follow the formation and evolution of clusters and their host galaxies in a cosmological context \citep{emosaicsI}, the fraction of massive clusters formed in mergers (with mass ratios $>$1:10) rises steeply from $\sim 10$ per cent in MW-mass haloes to $\sim 30$ per cent in haloes with mass $\Mhalo < 10^{11}\Msun$. This fraction increases to to $60{-}70$ per cent when considering GCs formed in mergers with mass ratios $>$1:100 for the same halo mass range. In the local Universe, starbursting dwarf galaxies (also known as blue compact dwarfs) display high gas surface densities and disturbed \HI kinematics with large velocity gradients which are indicative of a recent merger \citep{Lelli12a}. \citet{Lelli12b} note that starbursting dwarfs have more concentrated stellar and \HI distributions than typical dwarfs of the same mass. There is also observational evidence that star clusters with masses $\ga 10^5\Msun$ are currently forming in nearby starbursting dwarf galaxies \citep{Leroy18,Oey17,Turner17}. 

Despite their similar ages, the observed $\sim 0.5$ dex lower metallicity of the GCs relative to the stellar body of DF2 \citep{Fensch19} also seems to support the merger hypothesis. As a result of the mass-metallicity relation, a gas-rich satellite galaxy would contribute an important fraction of lower metallicity gas, diluting the composition of the clusters formed during the burst that follows the merger. The presence of a peak in the spectroscopically-determined star formation history of DF2's stellar body at $\sim 9\Gyr$ \citep{Ruiz-Lara19} further supports this scenario.

Following the merger and the formation of the GCs, violent relaxation will cause the old stars to settle in a configuration that is more extended than the original galaxies \citep{MovandenBoschWhite10}, similar to that of an elliptical galaxy. The relaxation process affects the orbits of all particles independent of their mass, such that the newly formed GCs are distributed uniformly in radius and avoid the rapid in-spiral that occurs in the starburst scenario examined in Section~\ref{sec:starburst}.

\subsubsection{Scenario III: Feedback-induced compression in gas-rich environment}

An additional possibility to obtain the high gas pressures necessary to form the massive GCs occurs in the gas-rich environments of high-redshift galaxies. \citet{Ma19} found that these conditions are common in simulations of a gas rich MW-mass halo at $z>5$. In their simulation, a $\sim \kpc$ size region is compressed by feedback from surrounding star-forming regions. The gas reaches very high densities and collapses gravitationally, while its own feedback is insufficient to stop the collapse and prevent the conversion of a large fraction of gas into stars to form a massive cluster with $M \ga 10^6\Msun$. The high gas fractions and star formation rates needed in this scenario are common at $z>1$, driven by the steep increase in the gas accretion rate onto galaxies \citep{Correa18}, and also in the major merger rate \citep[e.g.,][]{Fakhouri10}. This makes it likely that a major merger or gas accretion event could induce the initial star formation episode that triggers the gas compression required in this scenario. Unfortunately, separating these events is very difficult without additional information. Increasing the precision in the measurement of GC ages and metallicities would certainly improve the constraints. 

From this analysis we conclude that a merger-driven starburst is the most likely scenario, with the possibility of feedback-induced gas compression playing a role in producing the high gas pressures and shear predicted by the ICMF. Our model provides an upper limit to the fraction of stars formed in clusters during the GC formation episode (and therefore sharing the same metallicity). The difference in metallicity between the GCs and the stellar body suggests that a lower fraction of stars formed in bound clusters during the GC formation episode than the upper bound of 87 per cent we obtain here, and could indicate that DF2 formed some of its stars before or after the GCs. In the context of our results, this requires a narrower ICMF and more extreme ISM conditions, somewhere between the {\it constrained CFE} and the {\it minimum CFE} models (see Fig.~\ref{fig:DF2_conditions}).

\subsection{How did DF2 become so diffuse?}
\label{sec:expansion}

After the burst of star formation that produced the massive GCs, the galaxy must have lost its dense star-forming gas. Its high post-merger surface brightness, even higher than that of blue compact dwarfs, must have also decreased until it attained its present, diffuse morphology.
We may estimate the total energy required to unbind the dense stellar cusp in the merger remnant. For this we may use the virial theorem while assuming that stars dominate the mass, such that the potential energy is $W \approx \Mstar \vcirc^2$ \citep{Navarro96b}. Assuming the observed size of the most compact nearby dwarf galaxies with stellar masses $\sim 10^8\Msun$, $r_e \sim 100\pc$ \citep{Norris14}, and the DM circular velocity $\Vcirc(r=100\pc)\approx6.5\kms$ given by the NFW profile (Eq.~\ref{eq:NFW}) assuming the \citet{vanDokkum18a} halo mass and mean concentration\footnote{Assuming instead the halo mass expected from abundance matching, $\Mhalo = 10^{11}\Msun$, has a negligible effect on the binding energy.} (Eq.~\ref{eq:cm_relation}), we obtain $W = 4.1\times10^{52}\erg$. This binding energy amounts to less than 0.03 per cent of the total supernova energy produced by the observed GCs. Hence, it is an interesting possibility that the bursty and clustered SN feedback from the formation of several unusually massive GCs could have contributed to the transformation of the galaxy into a diffuse object through expansion. 

Assuming that the galaxy initially inhabited a typical DM content for its stellar mass \citep[$\Mhalo \sim 10^{11}\Msun$ as estimated using semi-empirical models e.g.,][]{Behroozi19, Moster18}, a substantial amount must have been lost from the central region in order to match the current upper limit on the dynamical mass from \citet{vanDokkum18a}, $\Mtot(r<3.1\kpc) < 3.2\times10^8\Msun$. Together, the reduction in the densities of {\it both} stars and dark matter in the central region of the galaxy are expected to be consequences of the process of feedback-driven DM core creation in dwarf galaxies \citep{Read16a, El-Badry16}. 

As discussed in Section~\ref{sec:conditions}, the {\it constrained CFE} model requires that a large fraction of the stars produced in the GC formation episode, $\approx 87$ per cent formed in bound clusters. Star formation that is highly clustered in space and time produces more energetic bursts of feedback, and these are more efficient at driving massive galactic-scale gas outflows \citep[e.g.,][]{Sharma14,Fielding17,Fielding18}. Massive outflows are known to drive the irreversible expansion of the central DM distribution via rapid oscillations of the potential \citep[][]{Navarro96b,ReadGilmore05,Mashchenko08,PontzenGovernato12,Read16a,Freundlich20}. Therefore, the high degree of clustering required to form the GCs in our model further supports a transformation scenario where stellar feedback from bursty star formation caused an irreversible expansion of the central DM distribution in DF2. Furthermore, as stars are also collisionless, their orbits would also expand, reducing the surface brightness of the galaxy. The formation of the inner DM core could also make the galaxy more susceptible to stripping. A recently proposed mechanism for the formation of UDGs through tidal stripping of normal galaxies relies on the presence of a shallow DM core to facilitate the stripping process \citep{Carleton19}.  Furthermore, the long-term survival of the GCs to in-spiral due to dynamical friction seems to disfavour a central DM cusp in dynamical models, and is consistent with a feedback-driven core transformation and expansion \citep{Chowdhury19}.

\begin{figure*}
    \includegraphics[width=\textwidth]{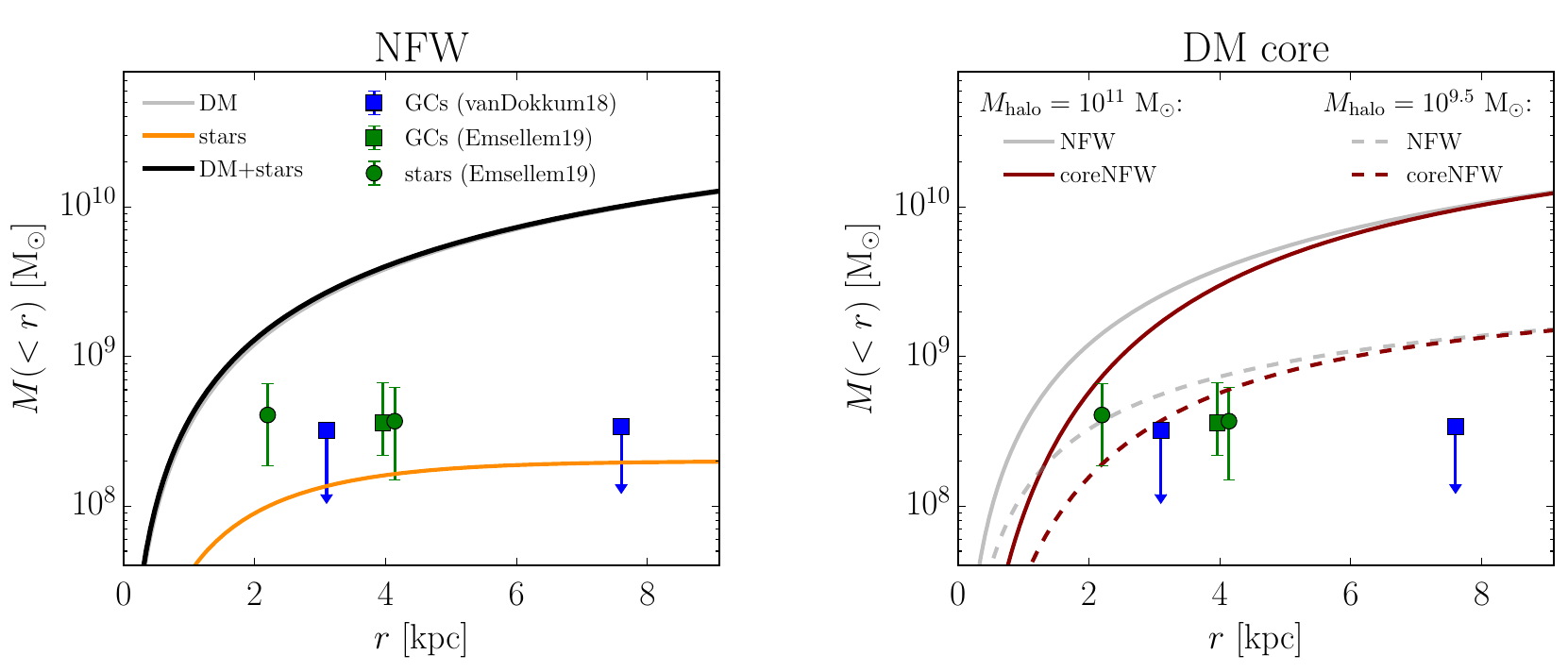}
    \caption{Estimates of the effect of feedback-driven core formation on the progenitor of DF2 at $z\ga 1.3$. Left: stellar, DM, and total mass profiles assuming the fiducial halo mass obtained from abundance matching, $\Mhalo = 10^{11}\Msun$. Right: DM-only profiles assuming the presence of a core using the coreNFW \citep{Read16a} profile. The red solid line shows the effect of the core for the fiducial mass, while the gray solid line reproduces the NFW profile from the left panel. Dashed lines show the same effect for the $\sim 30$ times lower halo mass needed to bring the density profile into agreement with observational constraints at $r\la 4\kpc$, $\Mhalo = 10^{9.5}\Msun$. Points with error bars in both panels indicate the dynamical constraints from the velocity dispersion of the GCs \citep[squares;][]{vanDokkum18a, Emsellem19}, and the stars \citep[circles;][]{Emsellem19}. Assuming the coreNFW model for the fiducial halo mass, the feedback-modified mass profile is inconsistent with constraints at all radii. Even assuming a 30 times lower halo mass, the enclosed DM mass at $r> 7\kpc$ is a factor of $\sim 10$ larger than the \citet{vanDokkum18a} constraint. }
\label{fig:core_profiles}
\end{figure*}

Figure~\ref{fig:core_profiles} estimates the effect of feedback-driven expansion on the mass profile of DF2. The left panel shows the mass profile for the fiducial halo mass $\Mhalo = 10^{11}\Msun$ (and mean concentration) obtained using the semi-empirical model by \citet{Behroozi19}. For the distance assumed here, all dynamical estimates (points with error bars) are inconsistent with this halo mass. The right panel shows the effect of a DM core based on the coreNFW model, which parametrizes the extent of the core using the projected stellar half-mass radius \citep{Read16a}, 
\begin{equation}
    M_{\rm cNFW}(<r) = M_{\rm NFW}(<r) f^n ,
\end{equation}
where
\begin{equation}
    f^n = \left[ \tanh{\left(\dfrac{r}{\rcore}\right)} \right]^n ,
\end{equation}
$n = 1$ (assuming a flat density core), and
\begin{equation}
    \rcore = 1.75 r_e ,
    \label{eq:rcore}
\end{equation}
and $r_e$ is the stellar half-mass radius. The right panel also shows the same result for a much lower host halo mass $\Mhalo/30 \approx 10^{9.5}\Msun$. This is the reduction in halo mass necessary to match the dynamical constraints at $r\la 4\kpc$ by \citet{vanDokkum18a} and \citet{Emsellem19}.

The amount of energy required to create the DM core can be estimated for a given value of $\rcore$ using the difference in binding energy between the coreNFW profile, and the initial pristine NFW halo,
\begin{equation}
    \Delta W = -\dfrac{1}{2}  \int_{0}^{\infty}{ \dfrac{G [ M_{\rm cNFW}(r)^2 - M_{\rm NFW}(r)^2 ] }{r^2}} \diff r .
\end{equation}
For the cored profile shown on the right panel of Figure~\ref{fig:core_profiles} (assuming the reduced mass $\Mhalo=10^{9.5}\Msun$), the formation of the $\sim 3.9\kpc$ core requires about 3 per cent of the total SNe energy produced by the 11 observed GCs alone (assuming $10^{51}\erg$ per supernova event). Using the value for the CFE $\sim 0.87$ obtained in Section~\ref{sec:GCMF}, we can calculate the total feedback contribution from all the clusters that formed $\ga 9\Gyr$ ago. The result shows that only $\sim 0.1$ per cent of the total SN energy produced by the clusters is necessary to create the core. This roughly matches the estimates for the SN energy used to transform cups into cores obtained by \citet{Penarrubia12} and \citet{Read16a}. However, the mass constraints at larger radii ($r > 7\kpc$) are more difficult to satisfy, even with the 30 times reduction in $\Mhalo$. This could indicate that feedback-driven expansion modifies the shape of the outer halo density profile, and highlights the need for detailed modelling of the effect of clustered feedback on the mass distribution of DF2.

\section{Where are the accreted GCs?}
\label{sec:accreted}

The major merger that we expect to have produced the observed GC population was likely due to a collision with a massive dwarf satellite galaxy of mass $\Mstar \sim 10^8\Msun$. Our model accounts for the GCs formed during the gas-rich merger, prompting the question of what happened to the GC population that each of the two progenitors may have formed prior to the merger. One possibility is that these GCs are part of a population of lower mass clusters which remain undetected in current observations due to the difficulty of removing background contamination. In fact, \citet{vanDokkum18b} find marginal evidence of detection of GCs with luminosities near the peak of the Milky Way luminosity function, $M_{V} \sim -7.5$. If a population of lower-mass GCs were to be discovered, these could correspond to the accreted clusters brought in by the progenitors, especially if there is a gap in the GCMF between the massive peak and the lower-mass population (i.e., the GCMF is bimodal)\footnote{After the submission of this manuscript, \citet{Shen21} extended the search for fainter GCs and found an additional $7.1^{+7.33}_{-4.34}$ GCs with luminosities $M_V > -8$, and a typical luminosity function. This newly discovered lower-mass GC population is separated from the massive peak by a gap in the total GCMF, which supports our conclusion that there was an episode of more quiescent star formation before or after (but plausibly the former) most of the massive GCs in DF2 formed.}. This scenario would be further supported if the low-mass GCs had metallicities that are clearly distinct from those of the massive GCs. The existence of this population would, however, not change the result of our analysis regarding the origin of the unusually massive GCs. It is in principle also possible that the progenitors did not host any GCs prior to the merger. However, their masses at $z\sim 1.3$, $\Mstar \approx 10^8\Msun$, are a factor of $\sim3$ above the value below which galaxies without GC systems are found in the local Universe, $\Mstar\sim 3\times10^7\Msun$ \citep{Forbes18b}. Had DF2's progenitors continued to grow normally, at $z=0$ their stellar masses would be large enough to ensure that they host at least one GC.

\section{Conclusions}
\label{sec:conclusions} 

We have modelled the effect of galactic environment on the formation of star clusters to reconstruct the properties of the progenitor of the ultra-diffuse galaxy NGC1052-DF2 at cosmic noon. Using the \citetalias{Trujillo-Gomez19} model for the ICMF and the \citet{Kruijssen12d} model for the bound fraction of star formation, which both reproduce the demographics of young cluster populations in the local Universe, we recovered the star forming conditions  $\ga 9\Gyr$ ago that could have led to the peculiar GC mass function DF2, which has a $\sim 4$ times higher median mass compared to the value found across nearly all galaxies.

Assuming an uncertainty range in the Toomre $Q$-parameter $0.5 < Q < 3.0$, our model shows that DF2's massive GCs  could have originated from a highly sheared and dense cold gas disc with $\SigmaISM \ga 10^3\Msunpc2$ and $\Omega \ga 0.7\Myr^{-1}$, similar to the conditions found in circumnuclear starbursts in the local Universe as well as in the Central Molecular Zone of the Milky Way. In addition, the model predicts that during the formation of the GCs, $\approx 87$ per cent of the stars in DF2 were born in bound star clusters. We have focused on the 11 GCs originally identified by \citet{vanDokkum18b}. Including the GC newly confirmed by \citet{Emsellem19}, which sits near the median luminosity of GCs used here \citep{Trujillo19}, would not change out results significantly. The slight increase in the height of the GCMF would require a further shift of the ICMF minimum cluster mass towards higher masses, requiring an environment with even higher gas density and shear.   

The extreme environmental conditions suggested by the GCMF could indicate either a progenitor with a high concentration of dense gas at its centre driven by internal dynamics (i.e. a nuclear starburst), or a gas-rich major merger event. In the first scenario, the gas surface density and shear values allow us to estimate that the central density of the galaxy must have been extremely high, with $\sim 4\times10^6\Msun$ of gas within the central region of galactocentric radius $r \la 30\pc$. However,  since this gas mass is not enough to form either the 11 GCs or the total mass under the ICMF (Figure \ref{fig:CMF_DF2}) at once, several separate bursts fed from a much larger gas reservoir would be required. Moreover, the formation of the GCs at such small galactocentric radii would have caused them to become the centre of the galaxy on an extremely short timescale. Except in the unlikely situation where each of the 11 GCs formed in its own progenitor galaxy and later accreted, the nuclear starburst scenario is ruled out by the observed absence of a nuclear star cluster in DF2. In the second scenario, a major merger induces strong tides and shocks which drive the ISM of the merging galaxies to very high pressures, while the large scale motions generate turbulent flows which reduce the effective Toomre mass. Indeed, in simulations that follow the formation of star cluster populations in a cosmological context, mergers account for a larger fraction of the massive clusters formed in galaxies with $\Mvir < 10^{11}\Msun$ compared to more massive galaxies. This scenario could keep the GCs afloat and avoid their rapid coalescence into a nuclear star cluster. The compression of a large amount of gas due to a surrounding burst of feedback during the merger likely contributed to creating the conditions necessary to form the massive GCs. The observed peak in the star formation history of the stellar body $\sim 9\Gyr$ ago \citep{Ruiz-Lara19} is consistent with the co-formation of the GCs and a significant fraction of the stars during a major merger. Interestingly, \citet{Lewis20} find marginal evidence of rotation in the GC system and its relative misalignment compared to the rotation of the stellar body. This misalignment could be further evidence of the merger. 

In the favoured merger scenario, the merger remnant has a high central stellar density, presenting the problem of how such a dense galaxy could have become so diffuse. We find that stellar feedback is a viable mechanism for transforming the dense merger remnant into the diffuse object observed at present with $r_e = 2.2\kpc$. The high recovered cluster formation efficiency indicates that star formation was very bursty and spatially concentrated, supporting the scenario where feedback energy injection could have caused the expansion of the stellar component of the galaxy. This leads to the interesting scenario in which it was the formation of DF2's extreme GC population that caused it to become a UDG.

However, the formation of a $\sim 4\kpc$ DM core predicted by the coreNFW model, while enough to expand the stellar component, still places the DM component well above the dynamical constraints on the total mass of DF2. A reduction of the DM halo mass by a factor of $\ga 30$ is required to fit the constraints at $r<4\kpc$, but still an order of magnitude above the \citet{vanDokkum18a} constraints at larger radii. We hypothesise that the scatter around the $\Mstar{-}\Mhalo$ relation at low galaxy masses could alleviate this tension. In a follow-up paper, we use a semi-empirical model to estimate the impact of scatter in the $\Mstar{-}\Mhalo$ relation on the clustering of star formation, stellar feedback, and the structural evolution of galaxies, and show that feedback from young GCs could cause UDGs to lose a large fraction of their inner DM \citep{Trujillo-Gomez21}. Ultimately, increasing the precision in the ages and metallicities of the individual GCs and the field stars should provide much more stringent constraints on the models and parameters considered here.

Recent studies have investigated the effect of tidal stripping on the inner DM density of satellites using numerical simulations \citep{Ogiya18,Maccio20}. In particular, \citet{Maccio20} simulated the effect of tidal stripping of a DM-dominated satellite with DF2's stellar mass due to a massive host. Starting with a dwarf galaxy with a cored DM profile formed in a cosmological simulation, they find that radial orbits with pericentres of a few kiloparsecs can reduce the DM density to the values inferred in DF2. Although stripping could explain the low DM content of DF2 and DF4, it does not explain their unusually massive GC populations. If the orbits of these galaxies are indeed eccentric enough, the DM stripping would be enhanced by the presence of a core carved by the intense feedback from the massive GCs, and these mechanisms together may explain the unusual properties of these galaxies.

In addition to the unusually large GC masses, \citet{vanDokkum18b} also find that the cluster radii are a factor of 2 larger than in the MW GCs. While our model does not explain the origin of the anomalous sizes, we note that GCs in M31 with masses $M>10^6\Msun$ have sizes $\sim 3-10\pc$, which is fully consistent with the sizes observed in DF2, and also consistent with a continuation of the mass-radius relation of young star clusters \citep[see fig. 9 in the review by][]{Krumholz19}. This implies that the GC sizes in DF2 are `normal', and only their mass function differs from that seen in other galaxies.

Lastly, we note that we have focused on the unusually bright peak of the GC luminosity function. Our analysis does not exclude the presence of fainter clusters whose formation  may have preceded the merger event\footnote{The $\sim 7$ new GCs confirmed by \citet{Shen21} after the submission of this manuscript have a normal luminosity function, and could correspond to the GC accreted from the progenitors of DF2 (see Section~\ref{sec:accreted})}. If the major progenitors of DF2 had masses $\Mstar \sim 10^8\Msun$ as required by a major merger, they would be massive enough to have brought their own GC populations. This could be verified if any fainter clusters had distinct ages or metallicities compared to the 11 massive GCs. Our analysis also accounts for the possibility that the majority of DF2's field stars are accreted. In this case, the merger starburst formed only the observed GCs, and the CFE during their formation is $\sim 100$ per cent.


\section*{Acknowledgements}

The authors would like to thank the anonymous referee for a critical and constructive review, and Pieter van Dokkum for helpful discussions about the manuscript. This work made use of the software packages {\sc numpy} \citep{numpy}, {\sc scipy} \citep{scipy}, and {\sc matplotlib} \citep{matplotlib}. STG, JMDK, BWK, and MRC gratefully acknowledge funding from the European Research Council (ERC) under the European Union's Horizon 2020 research and innovation programme via the ERC Starting Grant MUSTANG (grant agreement number 714907). JMDK gratefully acknowledges funding from the Deutsche Forschungsgemeinschaft (DFG, German Research Foundation) through an Emmy Noether Research Group (grant number KR4801/1-1) and the DFG Sachbeihilfe (grant number KR4801/2-1). MRC is supported by a Fellowship from the International Max Planck Research School for Astronomy and Cosmic Physics at the University of Heidelberg (IMPRS-HD).

\section*{Data availability}

The data underlying this article are available in the article.



\bibliographystyle{mnras}
\bibliography{merged}


\appendix

\section{Sensitivity of results to model assumptions}
\label{sec:uncertainties}

The \citetalias{Trujillo-Gomez19} model for the ICMF is based on the integrated feedback-regulated star formation efficiency in molecular clouds. As such, it assumes a constant value of the star formation efficiency per free-fall time of the cloud, $\epsff = 0.01$, with a systematic uncertainty of $\sim 0.5$ dex, based on the range of observationally determined values in the MW and nearby galaxies \citep{Krumholz19}. \citetalias{Trujillo-Gomez19} find that the predicted minimum truncation mass of the ICMF has a steep dependence on $\epsff$ for $\SigmaISM \ga 10^2\Msun$. To assess the effect of this uncertainty on the reconstruction of the galactic environment of the progenitor of DF2, we run the model assuming a broad range of values $\epsff=0.005{-}0.2$. This range is considerably larger than the uncertainty range estimated by \citetalias{Trujillo-Gomez19}, $\sim 0.5~\rm{dex}$,  but serves to illustrate the effect. 

\begin{figure*}
    \includegraphics[width=0.33\textwidth]{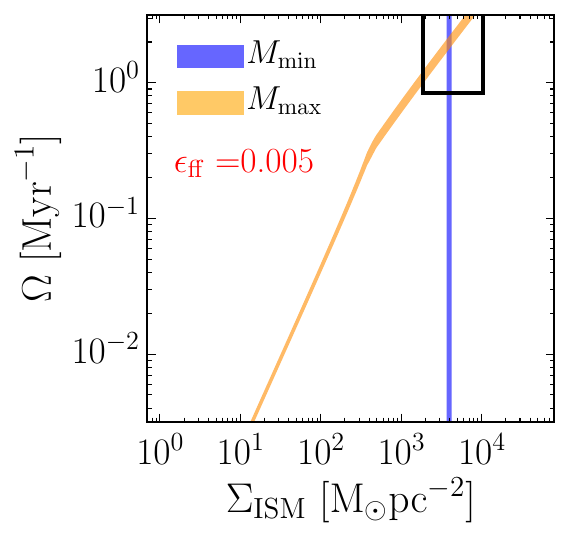}
    \includegraphics[width=0.33\textwidth]{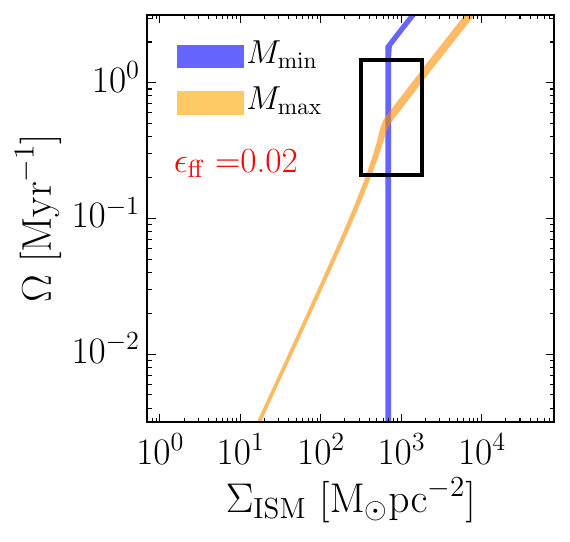}
    \includegraphics[width=0.33\textwidth]{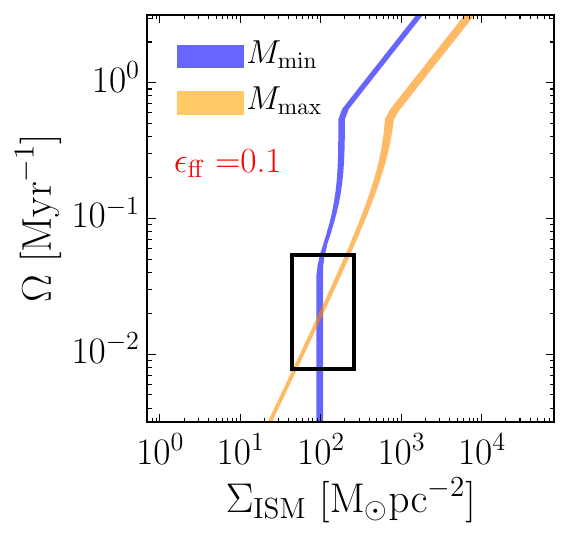}
    \includegraphics[width=\textwidth]{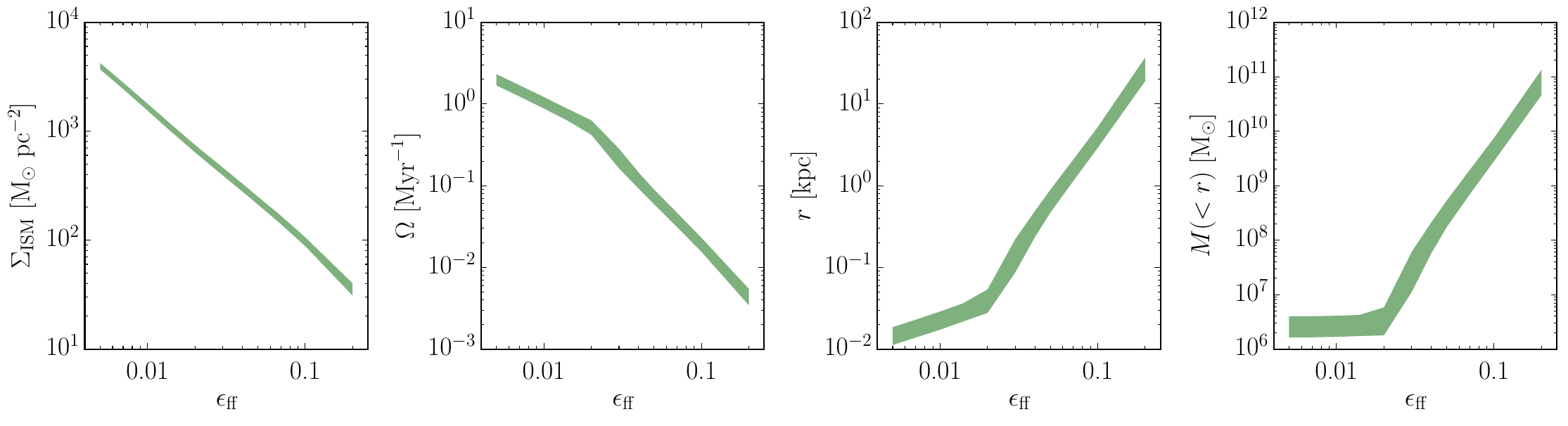}
    \caption{Effect of the uncertainty in the value of the star formation efficiency per free-fall time $\epsff$ assumed in the ICMF model. Top row: recovered DF2 environmental conditions for three values of $\epsff=0.005$, 0.02, and 0.1. The bottom row shows, from left to right, the dependence of the $\SigmaISM$, $\Omega$, the region radius, and the enclosed gas mass on $\epsff$. The shaded area in each panel indicates the full range covered by the physical conditions that produce the ICMF of DF2's GCs.}
\label{fig:DF2_epsff_test}
\end{figure*}

The top row of Figure~\ref{fig:DF2_epsff_test} shows the environmental conditions obtained using the {\it constrained CFE} model for three values of $\epsff$, assuming $Q=0.5$. The bottom row shows the dependence of the galactic environment on $\epsff$. The gas surface density and angular speed decrease steeply as $\epsff$ increases. 

This reduction in $\SigmaISM$ occurs because a larger star formation efficiency allows the bottom of the merger hierarchy to shift to lower cloud masses, decreasing the value of $\SigmaISM$ needed to obtain a given fixed $\Mmin$ \citepalias[see][]{Trujillo-Gomez19}. The effect of varying $\epsff$ on $\Mmax$ is relatively small in this regime, resulting in the reduction of the angular speed of the overlap solution region (where the minimum and maximum mass regions cross). As the gas surface density and angular speed decrease with increasing $\epsff$, the model allows for a larger size of the high density gas region where the massive GCs can form. The gas mass within the region increases beyond DF2's total mass $M \sim 3\times10^8\Msun$ for $\epsff > 0.06$, indicating that the required conditions become unphysical. For $0.03 \la \epsff \la 0.06$, the total gas mass of the nuclear starburst region becomes large enough to account for the total mass of clusters under the ICMF, and the formation radius large enough ($r \la 300\pc$) for the dynamical friction timescale to increase to $t_{\rm df} \la 100\Myr$. This means that if the GCs formed under these conditions, they would still have spiraled in by dynamical friction on a time-scale much shorter than their ages. As a result, $\epsff$ values in this range would overcome one of the challenges faced by the nuclear starburst scenario considered in Section~\ref{sec:starburst}. Such a scenario remains unfeasible since it requires fine-tuning of the environmental conditions to avoid a shift in the ICMF throughout the starburst region. Therefore, a variation in $\epsff$ would not change the qualitative conclusions of this work.

\section{Impact of distance uncertainty}
\label{sec:distance}

The distance to NGC1052-DF2 has been highly debated, with the two main estimates $D \approx 20\Mpc$  \citep[as assumed here following][]{vanDokkum18a}, and $D \approx 13\Mpc$ \citep{Trujillo19}. Interestingly, even though assuming the smaller distance would mean that DF2 is a much more typical dSph galaxy with $\Mstar = 8.5\times10^7\Msun$ and a typical GC mass function \citep{Trujillo19}, it would not change the high fraction of the galaxy stellar mass that is contained in its GCs, $\sim 5$ per cent. Most importantly, it would not change the fact that a very diffuse galaxy (with a half-light radius $r_e = 1.5\kpc$ for $D = 13\Mpc$) had in its past the high pressure conditions needed to form at least 11 GCs. Here we repeat the analysis of the origin of the DF2 GCMF using the lower proposed value of its distance in order to understand how the constraints on the evolution of the galaxy would change. For this, all the distance-dependent properties were scaled to $D=13\Mpc$, yielding a factor of $\sim 1.5$ reduction in the effective radii, and a factor of $\sim 2.4$ reduction in the galaxy and GC luminosities and stellar masses. 

\begin{figure*}
    \includegraphics[width=\columnwidth]{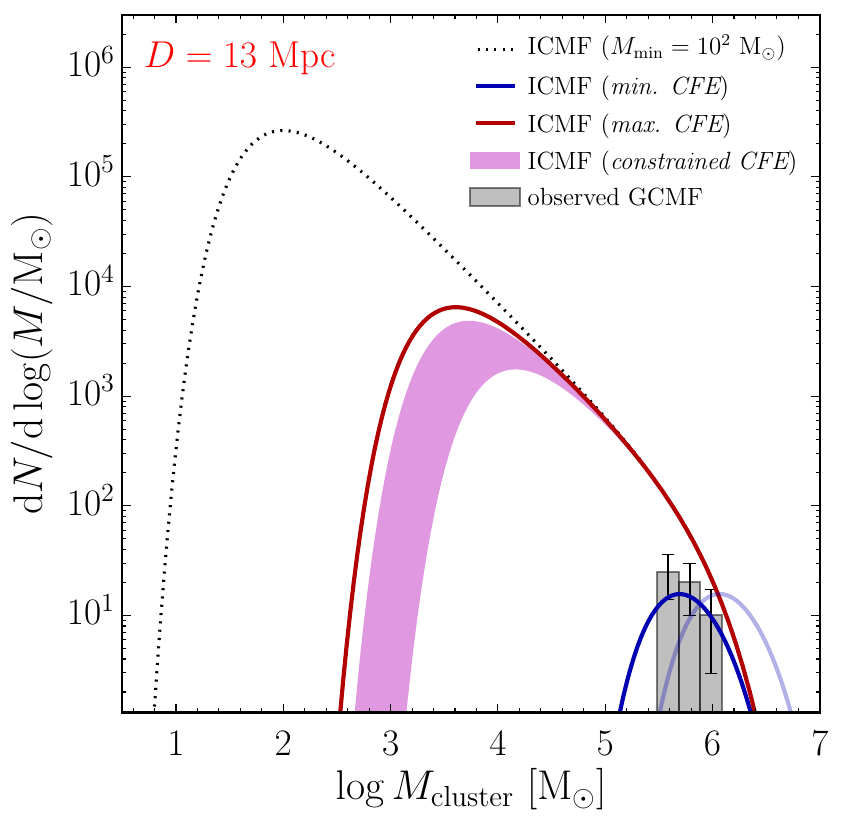}
    \includegraphics[width=\columnwidth]{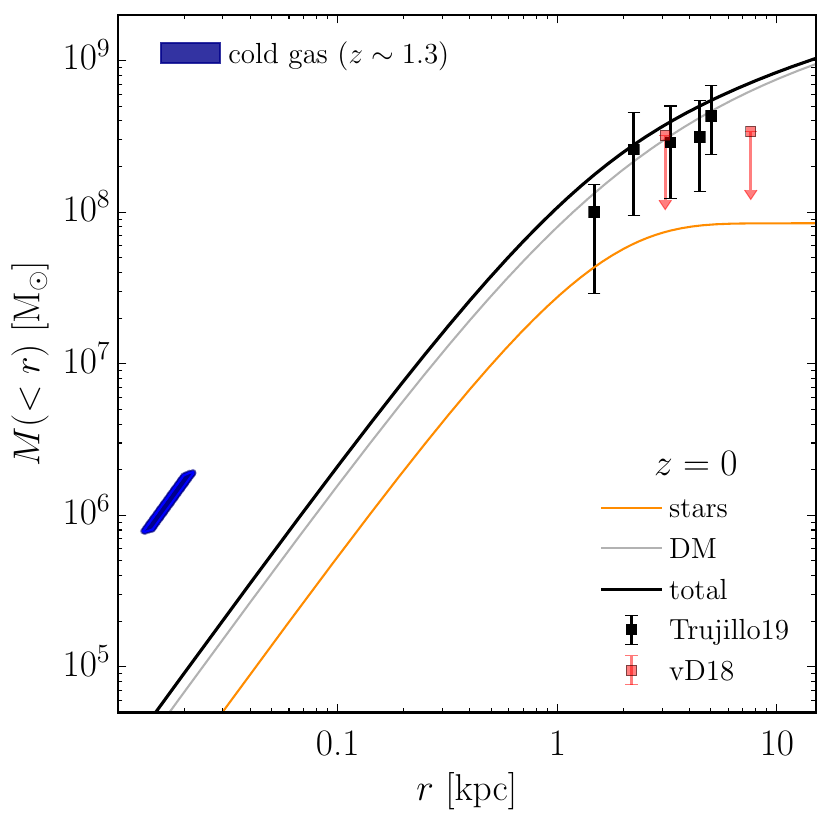}
    \caption{Effect of assuming a distance $D=13\Mpc$. Left: constraints on the ICMF of DF2's GCs. The fit to the GCMF for $D=20\Mpc$ is shown as a faint blue line for comparison. Right: density profile of star-forming gas at the time of formation of the GCs compared to present-day constraints (points with error bars) from \citet{Trujillo19} and \citet{vanDokkum18a}. Because both the galaxy and the GC masses are scaled down by the same factor to correct for the reduced distance, the fraction of mass in GC relative to the field stars remains unchanged relative to the \citet{vanDokkum18a} distance, while the minimum and maximum truncation masses of the ICMF are reduced by a factor of $\sim 2.4$. }
\label{fig:CMF_DF2_D13}
\end{figure*}

The stellar mass of DF2 assuming the smaller distance, $\Mstar = 8.5\times10^7\Msun$, is about twice the mass of the Fornax dSph (which hosts 5 GCs), $3.8\times10^7\Msun$ \citep{deBoerFraser16}. Figure~\ref{fig:CMF_DF2_D13} shows that, as expected, the ICMF simply scales down to lower cluster masses with respect to the ICMF assuming the original distance (see Figure~\ref{fig:CMF_DF2}). Because both the total stellar mass of DF2 and the total mass of its GCs scale by the same factor, the fraction of mass in GCs remains unchanged at $\sim 5$ per cent. Since the number of GCs is independent of distance, DF2's specific frequency would increase by a factor of 2.4. Since its mass is about twice the mass of Fornax and it contains about twice the number of GCs, DF2's specific frequency would be comparable to that of Fornax. Moreover, DF's GC system would be about twice as massive compared to that of Fornax, and hence the GC mass fraction in DF2 would also be comparable. Therefore, reducing the assumed distance to DF2 places it amongst the most extreme GC systems hosted by dwarf galaxies in the local Universe.   


\bsp	
\label{lastpage}
\end{document}